\documentclass[apj,numberedappendix]{emulateapj}
\usepackage[latin1]{inputenc}
\usepackage{amssymb}
\usepackage{amsmath}
\usepackage{graphicx}
\usepackage{xspace}
\usepackage{natbib}
\usepackage{url}
\usepackage{paralist}

\slugcomment{ }

\shorttitle{The jet-cycle in 3C120}
\shortauthors{Lohfink et al.}

\begin{document}

\title{An X-ray View of the Jet-Cycle in the Radio Loud AGN 3C120}


\author{Anne M. Lohfink\altaffilmark{1,2}, Christopher S. Reynolds\altaffilmark{1,2}, Svetlana G. Jorstad\altaffilmark{3,4}, Alan P. Marscher\altaffilmark{3}, Eric D. Miller\altaffilmark{5}, Hugh Aller\altaffilmark{6}, Margo F. Aller\altaffilmark{6}, Laura W. Brenneman\altaffilmark{7}, Andrew C. Fabian\altaffilmark{8}, Jon M. Miller\altaffilmark{6}, Richard F. Mushotzky\altaffilmark{1,2}, Michael A. Nowak\altaffilmark{5}, Francesco Tombesi\altaffilmark{1,2}} 

\altaffiltext{1}{Department of Astronomy, University of Maryland, College Park, MD 20742-2421, USA; alohfink@astro.umd.edu}
\altaffiltext{2}{Joint Space-Science Institute (JSI), University of Maryland, College Park, MD 20742-2421, USA}
\altaffiltext{3}{Institute for Astrophysical Research, Boston University, 725 Commonwealth Avenue, Boston, MA 02215, USA}
\altaffiltext{4}{Astronomical Institute, St. Petersburg State University, Universitetskij Pr. 28, Petrodvorets,  198504 St. Petersburg, Russia} 
\altaffiltext{5}{Massachusetts Institute of Technology, Kavli Institute for Astrophysics, Cambridge, MA 02139, USA} 
\altaffiltext{6}{Department of Astronomy, University of Michigan, 830 Dennison Bldg., 500 Church St., Ann Arbor, MI 48109-1042}
\altaffiltext{7}{Harvard-Smithsonian Center for Astrophysics, 60 Garden Street, Cambridge, MA, USA}
\altaffiltext{8}{Institute of Astronomy, University of Cambridge, Madingley Road, Cambridge CB3 0HA, UK}

\begin{abstract}
We present a study of the central engine in the broad-line radio galaxy 3C120 using a multi-epoch analysis of a deep \textit{XMM-Newton} observation and two deep \textit{Suzaku} pointings (in 2012). In order to place our spectral data into the context of the disk-disruption/jet-ejection cycles displayed by this object, we monitor the source in the UV/X-ray bands, and in the radio band. We find three statistically acceptable spectral models, a disk-reflection model, a jet-model and a jet+disk model. Despite being good descriptions of the data, the disk-reflection model violates the radio constraints on the inclination, and the jet-model has a fine-tuning problem, requiring a jet contribution exceeding that expected. Thus, we argue for a composite jet+disk model. Within the context of this model, we verify the basic predictions of the jet-cycle paradigm, finding a truncated/refilling disk during the \textit{Suzaku} observations and a complete disk extending down to the innermost stable circular orbit (ISCO) during the \textit{XMM-Newton} observation. The idea of a refilling disk is further supported by the detection of the ejection of a new jet knot approximately one month after the \textit{Suzaku} pointings. We also discover a step-like event in one of the \textit{Suzaku} pointings in which the soft band lags the hard band. We suggest that we are witnessing the propagation of a disturbance from the disk into the jet on a timescale set by the magnetic field.
\end{abstract}



\section{Generation of jets in Broad Line Radio Galaxies -- The jet cycle}\label{intro}

One of the most enduring questions surrounding active galactic nuclei (AGN) is the origin of powerful, relativistic, radio-emitting jets in some AGN. A closely related question concerns the extent to which we can draw an analogy between jets from stellar-mass black hole systems (particularly the Galactic microquasars) and AGN. In recent years it has become clear that many aspects of black hole accretion and jet formation are directly comparable between AGN and the lower-mass (about 10\,$M_\odot$) black holes in X-ray binary systems (XRBs) \citep{merloni:03a,mchardy:06a,fender:07a}. This is to be expected given the very simple scalings with mass for black holes in general relativity, although there is likely to be a larger diversity of environments and fueling mechanisms in AGN as compared with XRBs. XRBs can be observed as they cycle (secularly) through different accretion states, and we find that the properties and even the existence of the jets are closely tied to the accretion state (characterized by the spectral and timing properties of the X-ray emission) \citep{belloni:10a}. Importantly, it has been shown that XRB jets manifest themselves in two types depending upon the spectral state of the XRB. The first kind of jet is seen in the hard (corona dominated) spectral state as a continuous low-power outflow \citep{remillard:06a}. The second type of jet, which is similar to the jets seen in broad line radio galaxies (BLRGs), is launched when a low-mass XRB undergoes a transient outburst \citep{fender:04a}. The jet is usually observed close to the time of outburst maximum, as the source moves from the hard state to the soft (disk dominated) state \citep{fender:09a, millerjones:12a}. Low-mass black hole XRBs displaying these powerful jets are called microquasars \citep{mirabel:99a}. 

Assuming that these transient XRB jets and those from BLRGs are fundamentally the same phenomenon, the study of jets in AGN and XRBs give us different and complementary views of the physics. One major result learned from studying XRB jets is that the jet power, as well as the radiative efficiency of the accretion flow, can change dramatically in the same source at the same overall radiative luminosity on timescales far shorter than those associated with significantly changing black hole mass or angular momentum \citep{corbel:13a}. On the other hand, the much longer timescales displayed by AGN allow us to follow the complex relationships between individual jet-ejection events and the accretion disk, something which is difficult in XRBs.

It is now recognized that both Galactic microquasars and luminous radio-loud AGN (RLAGN) display complex cycles. In BLRGs, as first found by \citet{marscher:02a} in 3C120, major jet ejection events are preceded by strong dips in the X-ray luminosity.  Since the X-ray emission in BLRGs is thought to be dominated by the corona of the inner accretion disk, as shown for 3C120 by \citet{marshall:09a}, this X-ray/radio connection conclusively demonstrates a link between changes in accretion disk structure and powerful ejection events.  This behavior, which has parallels in the phenomenology of Galactic microquasars is one of the few observational clues that we have to the origin of radio jets. For example, GRS~1915+105 has been found to launch a relativistic jet knot as the inner-edge of the accretion disk moves all the way to the ISCO \citep{fender:04a}. This general idea has been confirmed in much more extensive monitoring campaigns of 3C120 and 3C111 by \citet{chatterjee:09a,chatterjee:11a} who also show that an observed correlation between the X-ray and optical flux can be explained in terms of the (inward or outward) radial propagation of powerful disturbances within the accretion disk \citep{mckinney:09a}. We note that \citet{tombesi:11a, tombesi:12a} recently reported a possible disk-outflow connection from a \textit{Suzaku} spectral analysis of 3C 111 and compared this to events in the radio jet.  

\begin{figure}[t]
\includegraphics[width=\columnwidth]{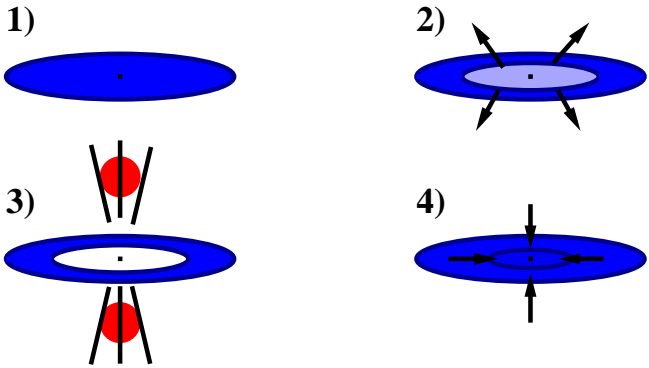}
\caption{Summary of our current understanding of the four steps of the jet cycle; 1) the accretion disk is full, 2) inner disk becomes instable, 3) jet is formed, 4) disk refills. A more detailed description of the jet cycle can be found in the text.}\label{jet}
\end{figure}

On the basis of these BLRGs studies, the current picture is that the jet and disk are linked via a ``jet cycle", summarized in cartoon form in Fig.~\ref{jet}.  The cycle starts with a full accretion disk extending all the way to the ISCO (Step 1).  Some instability, possibly associated with a breakdown in the inertial confinement of the magnetic flux bundle threading the black hole \citep{reynolds:06a,mckinney:12a}, destroys the innermost regions of the disk and ejects matter out of the disk plane (Step 2), causing an X-ray dip.  This triggers a powerful relativistic mass/energy ejection along the jet which, with a time delay corresponding to propagation and optical depth effects, results in a radio flare and the creation of a new VLBA knot/shock (Step 3).  One possibility is that this ejection event is triggered by the re-accretion of open magnetic field lines onto the rotating event horizon.   Lastly, the inner accretion disk refills (Step 4) and the cycle repeats. 

Relativistic X-ray reflection and, in particular, the broad iron line gives us a powerful tool to study the inner regions of AGN accretion disks \citep{tanaka:95a,fabian:95a,reynolds:03a,reynolds:12a}.  If our hypothesis of the jet-cycle in BLRGs is correct we would expect to see relativistic disk reflection during periods when the inner accretion disk is filled. When the inner disk is destroyed, we should see indications for a truncation of the disk.  Confirming this prediction would provide a major boost to the current jet formation scenario.  Measuring the inner radius of the accretion disk during several phases of the jet cycle will help to anchor this scenario.  

In this work we study 3C120 ($z=0.033$), a nearby and X-ray bright BLRG with a bulge dominated host, showing signs of a recent merger \citep{garcia:05a}. It is known to harbor a black hole with a mass of $(5.7\pm2.7)\times10^7\,\rm{M}_\odot$ \citep{pozo:12a}. 3C120 has been studied in detail at many wavelengths \citep[e.g.,][]{chatterjee:09a} and is well known to show a one sided superluminal jet.   It also exhibits a prominent jet cycle \citep{marscher:02a}, that has been previously studied multiple times in the X-ray range. Earlier \textit{XMM} observations analyzed by \citet{ballantyne:04a} and \citet{ogle:05a} find that the X-ray spectrum above 3\,keV can be well described by a power law and cold reflection, when accounting for neutral intrinsic absorption in the object. At soft energies they found a soft excess, which can be described by bremsstrahlung, or a power law. \citet{kataoka:07a} used \textit{Suzaku} data from 2006 and found indications for relativistic disk emission in the X-ray spectrum. These are confirmed by \citet{cowperthwaite:12a}, who find evidence for a truncation of the accretion disk in a re-analysis of the same dataset with updated models and calibration. 

From radio observations of superluminal motion and modeling of the recollimation shock, the jet angle has been estimated to about 16\,degrees \citep{agudo:12a}. With the caveat that the truncation of the thin disk may influence the ability of GR torques to align the inner disk, we assume a Bardeen-Petterson alignment of the jet and the inner accretion disk \citep{bardeen:75a} implying that the jet inclination can also be taken as the inner disk inclination. This, together with the fact that the UV/optical flux seems to mostly originate from the accretion disk \citep{ogle:05a}, makes 3C120 very suitable for studying potential changes in the accretion disk as expected from a disk-jet connection. It is also important to note that 3C120 was detected with \textit{Fermi}, allowing a tentative decomposition of the radio-to-$\gamma$ spectral energy distribution (SED) into jet and disk components \citep{kataoka:11a}.  This very simple decomposition suggests that the jet completely dominates the radio and $\gamma$-ray emission as expected, but is only a $\sim 10\%$ contributor to the optical, UV and X-ray emission. Given that the tentative decomposition suggests that 10\,\% of the X-ray flux comes from the jet, the much more careful treatment we plan to undertake is needed to determine whether the jet's flux can be ignored in fitting the X-ray spectrum or not.

In this paper, we test the jet-cycle picture using a detailed spectral analysis in the X-ray band supported by a multi-wavelength analysis in the optical/UV and radio.  A multi-epoch analysis of {\it XMM-Newton} and {\it Suzaku} spectra shows that, judged on purely statistical grounds, there are degenerate phenomenological interpretations of the X-ray spectrum, one of which does not require any relativistic disk reflection component.  However, physical considerations lead us to prefer a model in which the soft X-ray excess is a mixture of (steep) jet-emission and blurred ionized disk reflection.  Within the context of this model, we do indeed find evidence for changes in the inner accretion disk structure in the sense expected from the jet-cycle picture.   Additionally, we investigate the X-ray spectral variability on the timescale of a few hours, enabling us to tap into timescales comparable to the orbital time of the inner accretion disk and unobservable in X-ray binaries.  We suggest that the observed rapid X-ray spectral variability corresponds to magnetically mediated disturbances propagating from the disk (and/or disk corona) into the jet flow.

The paper is organized as follows: First, we describe the datasets used in this work and briefly discuss the data reduction techniques (\S \ref{data}). After placing the observations in the context of the jet cycle (\S \ref{context}) and performing a preliminary, basic investigation of spectral shape of the individual data epochs, a multi-epoch analysis is used to investigate the nature of the X-ray spectrum (\S \ref{spectral}). We then investigate the short-term variability and spectral energy distribution to further enhance our understanding of the processes driving the production of X-ray radiation in this AGN.  We end with a brief discussion of the implications for jet formation that can be drawn from the observations (\S \ref{dis}).  Throughout this paper, luminosities and distances are calculated using a $\Lambda$CDM cosmological model with $H_0=71\,{\rm km}\,{\rm s}^{-1}\,{\rm Mpc}^{-1}$, $\Omega_\Lambda=0.73$ and $\Omega_M=0.27$ \citep{komatsu:11a}.  For a redshift of $z=0.033$, this results in a luminosity distance to 3C120 of $143\,{\rm Mpc}$ and an angular size distance of $134\,{\rm Mpc}$.  

\section{Observations \& Data Reduction}\label{data}

The goal of this paper is a detailed analysis of the physical processes operating in the central engine of 3C120. To obtain this complete picture we conduct sensitive spectroscopy with \textit{XMM-Newton} and \textit{Suzaku}, which is placed in context by \textit{RXTE}, \textit{Swift} and VLBA monitoring data.  Below we describe the datasets that are considered and how they were reduced.   

\begin{table}[t]
\begin{center}
\caption{Overview of observations and exposures.\label{tab1}.   Superscript$^{\rm a}$ denotes that XIS0/XIS3 were used.   All {\it Suzaku} data were obtained after the failure of XIS2.}
\begin{tabular}{c c c c c}
\hline
\hline
Observatory & Instrument & Date & ObsID & Exposure  \\
&  &  &  & [ksec] \\ 
\tableline
\textit{Suzaku} (B) & XIS\tablenotemark{a} & 2012/02/14 & 706042020	 & 181.4 \\
\textit{Suzaku} (A) & XIS\tablenotemark{a} & 2012/02/09 & 706042010	 & 280.2 \\
\textit{XMM-Newton} & EPIC-pn & 2003/08/26 & 0152840101 & 89.2 \\
 \tableline
\end{tabular}
\end{center}
\end{table}

\subsection{Data Reduction}

\subsubsection{Suzaku \& XMM}\label{suz_red}
 
The datasets analyzed in this paper are from deep pointings by {\it XMM-Newton} and {\it Suzaku}. The \textit{XMM} data have already been analyzed by \citet{ballantyne:04a} and \citet{ogle:05a}. A summary of the observations (re-)analyzed here is presented in Table~\ref{tab1}. The two {\it Suzaku} pointings were taken as part of the {\it Suzaku AGN Spin Survey}, a cycle 4--6 Key Project (PI Reynolds).  

The Suzaku data were reduced with HEASOFT v6.12 and calibration files dated 2009 September 25. During our observations, XIS was operated in full-window mode. In preparing the XIS spectra, we first reprocessed and screened the data using \texttt{aepipeline} and the standard screening criteria, as listed in the \textit{Suzaku} ABC Guide. We then created individual spectra using \texttt{xselect} and response files using \texttt{xisrmfgen} and \texttt{xissimarfgen} tools for each detector and data mode combination. The regions used for source and background are circular with radii of 3.8' and 1.7' respectively for Suzaku A and 4.2' and 1.6' for the Suzaku B observation. Later the spectra of different data modes were combined for each XIS. While this is the recommended standard procedure to create XIS spectra, one additional complication needs to be accounted for in our case. XIS spectra are contaminated by absorption from a hydrocarbon layer residing on the optical blocking filter and we must take measures to correct for this contamination.  The contamination is monitored by the {\it Suzaku} team, and a correction is part of the construction of the standard effective area file.  At the time of our analysis, this contamination model was in the process of being updated due to inconsistencies with recent XIS observations.  We were able to use a test version of the contamination model in our analysis; for observations after 2010, this version is identical to the model incorporated in the CALDB files ae\_xi?\_contami\_20120719.fits, released by the XIS team as part of the 20120902 CALDB release\footnote{http://heasarc.gsfc.nasa.gov/docs/suzaku/analysis/xis\_contami2012.html}.  

For the \textit{Suzaku data} considered in this work, we found that we were unable to utilize data from the PIN detector due to the increased thermal noise caused by an increase in the leakage current.  While the problem is most apparent for temperatures above -11\,C, we already notice the effect (in terms of a distortion of the lower energy portion of the PIN spectrum) at temperatures of about -14\,C, the PIN temperature during our observation.   

The \textit{XMM-Newton} dataset was reduced using XMMSAS 11.0.0. The EPIC data were first reprocessed, using the calibration files as of February 2012. The spectra were then extracted using the tool \texttt{evselect}, selecting the default grade pattern. Source spectra are taken from a circular region with radius 41.2" centered on the source and the background spectra from an also circular region on the same chip with radius 28.9". The effective area files were generated using the XMMSAS task \texttt{arfgen}, and the redistribution matrices were produced using the task \texttt{rmfgen}. The EPIC detectors were operated in small window mode during the observation to reduce possible pile-up. During the observation, MOS1 was operated in timing mode. MOS2 alone has only $\sim$1/3 the effective area of EPIC-pn in the iron K region. Assessing possible pile-up effects using the single, double, triple and quadruple event pattern versus the predicted fractions utilizing \texttt{epatplot}, shows that EPIC-pn is not affected by pile-up while EPIC-MOS is influenced by pile-up. Considering these facts, we only consider EPIC-pn data for the analysis performed in this paper, as MOS2 would not significantly contribute to any constraints. The average observed net pn and MOS source count rates are 14.2\,cts$\,\textbf{s}^{-1}$ and 5.8\,cts$\,\textbf{s}^{-1}$ respectively. 

For the spectral analysis, all XIS/EPIC-pn spectra were binned to a signal-to-noise ratio of 10. The considered energy ranges are 0.7-1.7\,keV and 2.3-10\,keV for \textit{Suzaku}-XIS, with the energies around 2\,keV being excluded because of known calibration issues around the mirror (gold) and detector (silicon) edges. The EPIC-pn spectrum is used in the 0.5-10\,keV band. We extended all energy grids to energies far beyond the upper energy limit given by the highest data bin considered in fitting, in order to enable a proper model evaluation of the relativistic blurring.   

In any detailed spectral analysis such as is presented in this paper, we must be cognizant of the possibility of instrumental calibration errors.  Such errors would directly lead to systematic errors in our spectral fits.  The absolute flux calibration of X-ray observatories can be uncertain to $\sim 10\%$, but {\it our} physical conclusions derive purely from the {\it shape} of the spectrum and hence we need only be concerned with the relative calibration.   As already mentioned, our {\it Suzaku} spectra are clearly inflicted by significant unmodeled calibration features around the Silicon K-edge and the Gold M-edge; we remove these energies from consideration by masking data in the 1.7--2.3\,keV band.   Our choice of low-energy cutoffs (0.5\,keV and 0.7\,keV in {\it XMM-Newton} and {\it Suzaku} respectively) is also driven by the need to exclude more poorly calibrated regions of the spectrum.   However, due to the lack of a practicable approach for propagating calibration uncertainties into the systematic error budget for our complex (large parameter space) spectral models, no further attempt is made in this paper to account for calibration uncertainties.    We note that the relevant calibration errors are thought to be small --- at the time of writing, the \textit{XMM}/EPIC has a maximum error in the relative effective area of 5\,\% (with an rms error that is rather smaller) and a maximum error in the absolute energy scale of 10\,eV \footnote{http://xmm.vilspa.esa.es/docs/documents/CAL-TN-0018.pdf}; for \textit{Suzaku} the maximum error on the energy scale is also 10\,eV and the effective area uncertainty is small outside of the mirror edges \footnote{http://web.mit.edu/iachec/meetings/2012/Presentations/Miller.pdf}.   

\subsubsection{RXTE}

A primary goal of this paper is to investigate changes in accretion disk structure during the jet-cycle in 3C120.  Hence, we require monitoring data to diagnose the state of the source at the time of the deep spectroscopic observations by {\it XMM-Newton} and {\it Suzaku}.  For the period 2002--2008, this monitoring was provided by {\it RXTE}.  The average cadence of the monitoring over this period is 4\,days, with the exception of a 2-month period each Spring when 3C120 lies too close to the Sun.  We have obtained these data from the HEASARC archives and reduced them exactly as described in \citet{chatterjee:09a}.   Only data from the top layer of proportional counter unit 2 were used.  For each pointing, we produce a background-subtracted PCA spectrum using the appropriate epoch-dependent background/response files. The fluxes quoted below were obtained from the spectra for each pointing, assuming a simple power-law continuum with a neutral absorption column of $1.23\,\times\,10^{21}\,$cm$^{-2}$ \citep{elvis:89a}. We do not consider any {\it RXTE} intra-pointing variability in this work. The resulting {\it RXTE} lightcurve is shown in Fig.~\ref{rxte_all}.

\subsubsection{Swift}

Monitoring from {\it RXTE} is not available after 2008.   Thus, we established a monitoring campaign with {\it Swift} to support the 2012 {\it Suzaku} pointings. XRT and UVOT data from this campaign are used in this paper.  The source can achieve X-ray flux levels that would lead to significant photon piled up in the XRT if operated in photon counting mode.  To reduce the impact of pile-up we preferred to use the windowed timing mode data when available and otherwise excluded the central region of the PSF to mitigate pile-up. The  \textit{Swift}-XRT data were first reprocessed to apply the newest calibration (XRT Calibration Files: 20120830). From the resulting event file, a spectrum was extracted using XSELECT following a standard extraction for the windowed timing mode as described in the ``XRT User's Guide"\footnote{http://heasarc.nasa.gov/docs/swift/analysis/xrt\_swguide\_v1\_2.pdf}. To properly account for the rolling of the satellite during the pointings (especially important in windowed timing mode), a spectrum is reduced for each good time interval with a different roll angle. Finally, these good time interval spectra are summed to yield the final spectrum for each pointing. As with the {\it RXTE} monitoring, X-ray fluxes are derived on the basis of a simple power-law continuum with a neutral absorption column of $1.23\,\times\,10^{21}\,$cm$^{-2}$.

The UVOT analysis is based on level II products. Each individual UVOT filter data file contains, in general, a number of exposures which were summed using the tool UVOTIMSUM. The UVOTSOURCE tool was then used to extract magnitudes from simple aperture photometry. Source and background regions were created for this purpose, with the position of the source region being obtained from the NASA Extragalactic Database (NED) in the first instance. The required source region and background region where both circular with a 4.8 arcsec radius. 

The resulting UV and X-ray lightcurves from {\it Swift} are shown in Fig.~\ref{swift_uv}.

\subsection{VLBA}

\begin{figure}[!hb]
\begin{center}
\includegraphics[width=0.40\columnwidth]{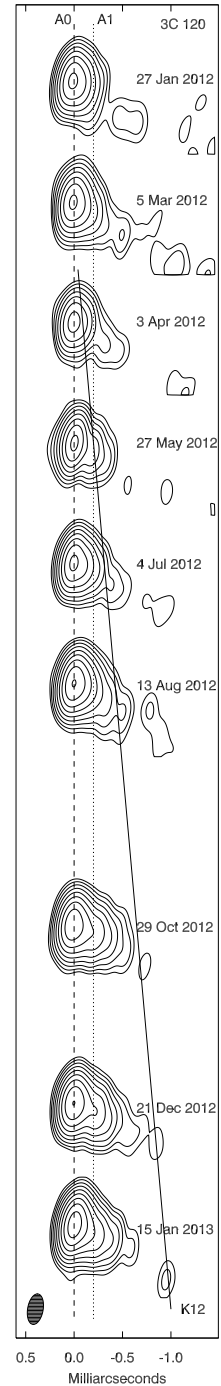}
\end{center}
\caption{Time sequence of VLBA images at 43 GHz of 3C120 at nine epochs. Contours correspond to 0.5, 1, 2, ..., 64, and 90\% of the maximum intensity of 1.8 Jy beam$^{-1}$ (reached on 2013 January 15). The easternmost feature, the ``core,'' is presumed to be stationary.
Knot {\it K12} is marked, with a mean trajectory (shown by line) correponding to an
apparent speed of and an ``ejection'' date (when the brightness centroid
of {\it K12} coincided with that of the core) of 2012 March $15\pm 10$. The elliptical
Gaussian restoring beam of FWHM dimensions $0.34\times 0.14$ milliarcsec along position
angle $-10^\circ$ is displayed in the bottom left corner.}
\label{vlba}
\end{figure}

We observed 3C120 nine times with the Very Long Baseline Array (VLBA) between 2012 January 27 and 2013 January 15 at a frequency of 43 GHz. The observations, part of the Boston University $\gamma$-ray blazar monitoring program (www.bu.edu/blazars) included ten 3-5 minutes scans
on 3C120 at each epoch.  The raw data were recorded at each antenna and correlated at the National
Radio Astronomy Observatory's Array Operations Center in Socorro, NM. The
resulting data on the {\it uv}-plane were edited, calibrated (including sky opacity
corrections and fringe fitting), and imaged in a standard manner with
routines from the Astronomical Image Processing System and Difmap software
packages. This involved an iterative procedure alternating imaging the
CLEAN algorithm with self-calibration of both phases and intensities,
which eventually converged to the final images.   The resulting images, with an angular resolution of 0.14 milliarcsec along the direction of the jet, are displayed in Figure \ref{vlba}. They reveal the appearance of a new superluminal knot, which we
designate {\it K12}, with a proper motion of $1.14\pm 0.22$ milliarcsec yr$^{-1}$, 
corresponding to $2.6\pm 0.5c$ at the distance of 3C120. The centroid of the
knot crossed the centroid of the ``core'' (the bright, presumed stationary feature at the
eastern end of the jet) on JD $2456001\pm 10$ (March, 14th 2012). Another development is the
enhanced brightness of the jet between 0.2 and 0.5 milliarcsec from the core that is
apparent starting with the 2012 August 13 image. This feature is either stationary or
moving subluminally.

\section{The Observations in Context}\label{context}

\begin{figure}[!hb]
\includegraphics[width=\columnwidth]{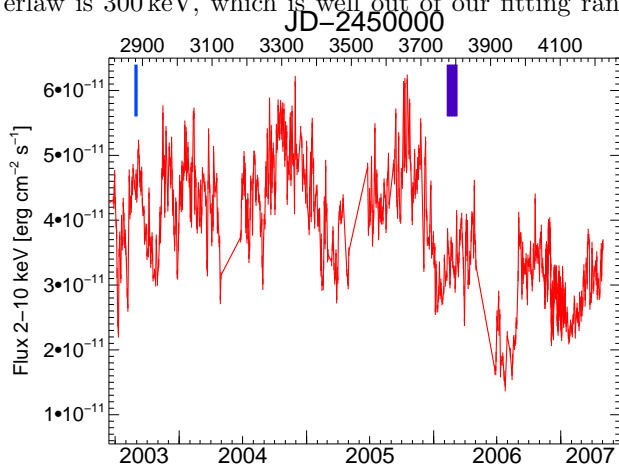}
\caption{Long term \textit{RXTE}-PCA monitoring for the 2--10\,keV flux, marked are the positions of the 2003 \textit{XMM} pointing and the position of the 2006 \textit{Suzaku} pointings.}\label{rxte_all}
\end{figure}

As outlined above, we need to estimate the placement of each of our spectroscopic observations in the jet-cycle. The long-term light curve taken by \textit{RXTE}, previously published by \citet{chatterjee:09a}, clearly shows the pattern of X-ray dips and recovery associated with the jet cycle (Fig.~\ref{rxte_all}).  We mark on Fig.~\ref{rxte_all} the time of the {\it XMM} pointing considered in this work.   This reveals that the pointing was on the peak of the flux sequence, corresponding, according to the model, to a radiatively-efficient disk extending down close to the black hole (Step 1 of Fig.~\ref{jet}).   Also marked is the position of the 2006-\textit{Suzaku} observation as analyzed by \citet{kataoka:07a} and \citet{cowperthwaite:12a} which occurred during a period of rising X-ray flux, as it was already re-analyzed recently we decide to not include it in our spectral analysis.   This rising flux may correspond to a refilling accretion disk (Step 4 of Fig.~\ref{jet}), in line with the truncated accretion disk ($r_{\rm in}\sim 10R_g$) found by \citet{cowperthwaite:12a}.

\begin{figure}[!ht]
\includegraphics[width=\columnwidth]{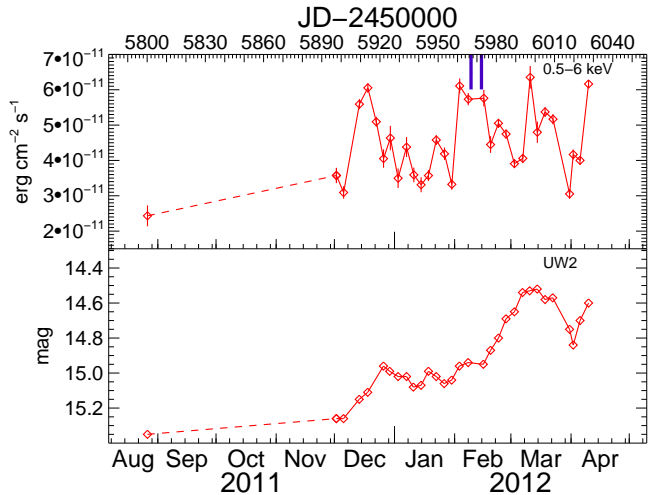}
\caption{\textit{Swift} monitoring of 3C120: X-ray and UV (UVW2 filter) results. The markers point out the beginning of the two parts of the deep \textit{Suzaku} stare.}\label{swift_uv}
\end{figure}

The \textit{Swift} monitoring campaign supported the new 2012-{\it Suzaku} pointings.   The monitoring (Fig~\ref{swift_uv}) shows that the general trend in the UV band was a rising flux, with some temporary stalling of the rise.  The X-ray variability is less coherent, showing indications for short-lived flux dips, but appears to show a rising trend. The X-ray flux is estimated by fitting an absorbed power law with the absorption column fixed to $3.1\times 10^{21}$\,cm$^{-2}$ (a value similar as the one found later for the best fit to the \textit{Suzaku} pointings). Similar short-timescale X-ray variability appears when zooming into the \textit{RXTE} X-ray monitoring; see Figure\,1 of \citet{marshall:09a}. 

The VLBA monitoring also helps to place the 2012-\textit{Suzaku} observations into context. The superluminal knot appears approximately one month after our \textit{Suzaku} pointing.  However, the work of \cite{chatterjee:09a} shows that there is approximately a 2 month delay between the probable time of the physical ejection from close to the black hole and the appearance of a new VLBA knot, most likely corresponding to the propagation time of the ejector between the black hole and the VLBA core.   Accounting for this time delay, we estimate that the actual ejection occurred approximately one month prior to the {\it Suzaku} pointings.  Hence, the prediction is that {\it Suzaku} should find a truncated/refilling accretion disk.  

In summary, on the basis of the {\it RXTE}, {\it Swift} and VLBA monitoring together with the jet-cycle hypothesis,  the 2003-\textit{XMM} pointing should show a filled accretion disk extending all the way to the ISCO. During the two \textit{Suzaku} pointings on the other hand we expect the accretion disk to be still truncated but be in a stage of refilling.

\section{Spectral Analysis}\label{spectral}

We begin by discussing our detailed analysis of the deep spectral {\it XMM-Newton} and {\it Suzaku} data. Our main technique is a  ``multi-epoch analysis" in which all spectral data are fitted together, tying together parameters that must (on physical grounds) be common across all epochs.  However, we begin by discussing an initial exploration of the spectra and the construction of the spectral models.  Throughout this work, the spectral analysis was performed with the Interactive Spectral Interpretation System\footnote{http://space.mit.edu/cxc/isis} \citep[ISIS Version 1.6.0-7;][]{houck:00a} using the newest \textsl{XSPEC} 12.0 models \citep{arnaud:96a}. All uncertainties are quoted at the 90\,\% confidence level for one parameter of interest ($\Delta\chi^2=2.7$). Systematic errors arising from the physical assumptions made in the models used for describing the spectra are not included in the given errors.

\subsection{Initial Data Exploration and Model Construction}\label{initial}

\begin{figure}[!ht]
\centerline{
\includegraphics[width=\columnwidth]{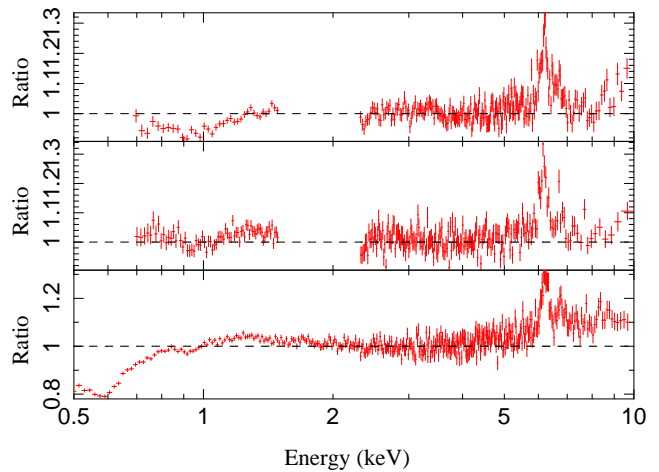}
}
\caption{Residuals to a simple absorbed power law (\textit{Suzaku} I [Top Panel], \textit{Suzaku} II [Middle Panel] and \textit{XMM} [Bottom Panel])  fitted from 2-4.5\,keV; spectra are rebinned for plotting.}\label{basic}
\end{figure}

Investigating the spectral variability by looking at the hardness evolution within each \textit{XMM} and \textit{Suzaku} pointing, we find it is mild, and therefore ignore it in this Section. The short-term variability that we do find will be discussed in \S\ref{var}. To get a first look at the spectral shape, we fit a simple absorbed power law to the 2--4.5\,keV data and then extend to the full energy range of the observation (Fig.~\ref{basic}). This reveals an iron line and an upturn at higher energies indicative of the presence of reflection in the spectra. Below 2\,keV, additional absorption beyond the Galactic column that is already accounted for is apparent.

Building upon this knowledge and previous works \citep{kataoka:07a,cowperthwaite:12a}, we construct three multi-component spectral models to describe our spectra.   Common to all of these models is a primary powerlaw continuum with photon index $\Gamma$ together with cold, distant reflection of that powerlaw described by the model \texttt{pexmon} \citep{nandra:07a}.  The reflection fraction $R$ and the iron abundance characterizing this reflection are left as free parameters.  As some parameters for \texttt{pexmon} cannot be determined from our fits we need to fix them to certain values.   We assume that the high-energy cutoff of the continuum powerlaw is 300\,keV, which is well out of our fitting range.   Since {\tt pexmon} strictly models reflection from a planar surface, we also need to fix the inclination parameter of the {\tt pexmon} model.   The most likely geometries for the material producing the cold/distant reflection features are either cloud-like or a large-scale torus, and thus a range of inclinations will contribute to the observed spectrum.  Assuming that this can be approximated as reflection from isotropically oriented planar segments, the most likely inclination angle is 60$^\circ$.  Thus, we set the inclination of the {\tt pexmon} component to 60$^\circ$.    We have verified that this assumption does not drive any of the conclusions reached in this paper --- in particular, repeating the analysis presented in \S\ref{jetdisk} using a {\tt pexmon} inclination of 30$^\circ$ yields best fitting parameters that are within the statistical error bars of our canonical analysis, with the exception of the reflection fraction.

Also common to all of our spectral models is a photoionized emitter producing, most notably, Fe XXV/Fe XXVI iron-K$\alpha$ lines.  The photoionized emission is parametrized by ionization parameter and normalization. It is described using a table model calculated using \texttt{xstar2xspec} from the \texttt{XSTAR} model \texttt{photemis}\footnote{http://heasarc.nasa.gov/docs/software/xstar/xstar.html}, assuming irradiation by a power law with $\Gamma=2$ and a number density of $10^{10}\,\text{cm}^{-3}$.  The {\tt xstar} model requires an assumed column density; we fix this to be $10^{22}$\,cm$^{-2}$, noting that there is only a very weak dependence of the spectral shape on this column density. As noted in \citet{lohfink:12b} there can be a strong degeneracy between Fe\,XXV and the blue wing of the broad iron line, therefore the inclusion of this component is crucial.  Finally, photoelectric Galactic absorption ($N_{H,Gal}=1.1\times 10^{21}\,{\rm cm}^{-2}$, the average value for 3C120 from HEASARC's nH Tool) is modeled with \texttt{TBnew}\footnote{http://pulsar.sternwarte.uni-erlangen.de/wilms/research/tbabs/} a newer version of \texttt{TBabs} \citep{wilms:00a} with cross sections set to \texttt{vern} and abundances set to \texttt{wilm}. We also account for intrinsic cold absorption in the source using \texttt{TBnew} with column density as a free parameter.

Our first spectral model is a ``disk reflection model".  To the above base model, we add a component describing relativistic reflection from the innermost regions of an ionized accretion disk.  To describe the ionized reflection associated with reflection of the inner accretion disk, we used a modified version of \texttt{reflionx} developed by \citet{ross:05a}, see \citet{lohfink:12b} for details.  The iron abundance is fixed to that characterizing the distant reflection (thus we assume a chemically homogeneous central engine; see \citealt{reynolds:12a}), and the photon index of the continuum irradiating the disk is tied to the one for the primary powerlaw.  The ionized reflection component is then relativistically blurred using the model \texttt{relconv} \citep{dauser:10a}; this naturally gives rise to a relativistic iron line, blurred Compton reflection hump and a soft excess. The radial dependence of the emissivity of the reflection component is assumed to have a power-law
  form with index $q$. The inner edge of the X-ray reflection regime can be at the ISCO \citep{reynolds:08a} or further out, and the outer edge was fixed to 400\,$R_\text{g}$.  Provided that $q>2$, the relativistic blurring kernel is only weakly dependent upon this outer radius. The accretion disk inclination $i$, and the black hole spin $a$ were left as free parameters.

For a radio-loud AGN relativistic disk reflection is not the only possible origin for the soft X-ray excess. A contribution from the jet emission could also produce an excess flux at soft energies.   Thus, our second spectral model is a ``jet model".  To the base model, we add a second (steep) powerlaw component with photon index $\Gamma_S$ to characterize a possible contribution from the jet.  As compared with the disk reflection model, this decouples the soft excess from the iron-K band structure.   This is motivated by similar treatments in \citet{kataoka:07a} and \citet{ballantyne:04a}.

The final spectral model is the ``disk$+$jet model" in which we include both a relativistic disk reflection component and a steep power-law.  

\subsection{Multi-Epoch Fitting}\label{me_fitting}

\subsubsection{Relativistic disk reflection model}

\begin{table*}[t]
\begin{center}
\caption{Spectral Parameters for 3C120 multi-epoch disk-reflection model with inner disk radius fixed at the ISCO; see the text for a detailed description of the model. \textit{Suzaku} spectra are normalized to XIS0 data. The power law normalization is $\text{photons}\,\text{keV}^{-1}\,\text{cm}^{-2}\,\text{s}^{-1}$ at 1\,keV. Bold values are the ones which are constrained by all observations as they are assumed to be constant.\label{mtable_base} 
}
\begin{tabular}{ccccc}
\hline 
& & Suzaku A & Suzaku B & XMM \\
 \hline \hline
 absorption & $N_\mathrm{H} [10^{22}$\,cm$^{-2}]$ & 0.59$_{-0.08}^{+0.05}$ & 0.41$_{-0.08}^{+0.11}$ & 0.74$_{-0.02}^{+0.01}$\\
continuum \& & $A_\text{pex}$ [$10^{-3}$] &  14.34$_{-0.08}^{+0.11}$ & 15.38$_{-0.09}^{+0.15}$ & 1.27$_{-0.05}^{+0.09}$ \\
cold reflection & $\Gamma$ & 1.85$_{-0.01}^{+0.01}$ & 1.92$_{-0.01}^{+0.01}$ & 1.50$_{-0.00}^{+0.01}$ \\
 & $R$ & 0.37$_{-0.06}^{+0.06}$ & 0.39$_{-0.08}^{+0.08}$ & 2$_{-0.33}^{+0}$\\
 \hline
ionized reflection &$A_\text{reflionx}$ [$10^{-4}$] & 1.19$_{-0.33}^{+0.34}$ & 1.47$_{-0.41}^{+0.40}$ & 2.87$_{-0.05}^{+0.07}$\\
 & $Fe/Solar$ & \multicolumn{3}{c}{\textbf{0.96}$_{-0.08}^{+0.13}$} \\
& $\xi$ [erg\,cm\,s$^{-1}$] & 10.0$_{-7.3}^{+7.7}$ & 10.0$_{-7.2}^{+3.7}$ & 5886.1$_{-85.3}^{+31.9}$\\
\hline
relativistic blurring & $q$ & 2.0$_{-0}^{+0.1}$ & 2.0$_{-0}^{+0.2}$ & 10$_{-5.4}^{+0}$\\
  &$a$ & \multicolumn{3}{c}{\textbf{$<0.29$}}\\
 & $i$ [deg] & \multicolumn{3}{c}{\textbf{60}$_{-2}^{+0}$}\\
 \hline
plasma & $A_\text{phot}$ & \multicolumn{3}{c}{\textbf{8.29}$_{-0.39}^{+38.85}$}
\\
 &$\log\xi$ & \multicolumn{3}{c}{\textbf{6.2}$_{-0.0}^{+0.8}$} \\ 
\hline
individual & $\chi^2$/dof & 1023.1/943 & 1199.6/1404 & 1844.1/1522 \\
 & $\chi^2$/dof & \multicolumn{3}{c}{4066.9/3875} \\
 & $\chi^2_\text{red}$ & \multicolumn{3}{c}{1.05} \\
 \hline\hline
\end{tabular}
\end{center}
\end{table*}

We fitted the disk reflection model to the {\it XMM-Newton} and {\it Suzaku} data employing multi-epoch fitting --- all spectra are fitted together, tying together the black hole spin, accretion disk inclination, and iron abundance that on physical grounds should be constant across epochs. The ionization state and normalization of the photoionized emitter are also tied across epochs since it is believed that this component originates from a spatially extended structure.   As a first approach, we assume that the inner radius of the accretion disk is fixed to the ISCO.  We find an acceptable fit overall($\chi^2/{\rm dof}=4067/3875$ [1.05]) and report the spectral parameters in Table~\ref{mtable_base} (ordered by the flux sequence established in \S\ref{context}).    As the source brightens in flux, the ionization state of the inner disk rises from $\xi\sim 10$ to $\xi\sim 5900$.  Along with the ionization state, the contribution of ionized reflection to the X-ray spectrum increases from 2\,\% to 85\,\% based on the unabsorbed model fluxes in 0.7-10\,keV. While, taken at face value, this makes the \textit{XMM} spectrum reflection dominated, the actually visible absorbed flux is still dominated by the cold reflection and power law continuum.  The accretion disk inclination determined from this fit is high and not consistent with the radio measurements ($\sim\,16$\,deg), possibly highlighting a mis-modeling of the data. Another problem is the tremendous difference in the normalization of the corona between the \textit{Suzaku} pointings and the \textit{XMM} pointing, a factor of 10 seems physically hard to explain. This in combination with the poor quality of the fit for the \textit{XMM} spectrum and the low spin value ($a<0.29$), could be an indication for a time variable recession of the accretion disk. 

\begin{table*}[t]
\begin{center}
\caption{Spectral Parameters for 3C120 multi-epoch disk-reflection model with free inner disk radius; see the text for a description of the model. \textit{Suzaku} Spectra are normalized to XIS0 data. The power law normalization is $\text{photons}\,\text{keV}^{-1}\,\text{cm}^{-2}\,\text{s}^{-1}$ at 1\,keV. Bold values are the ones which are constrained by all observations as they are assumed to be constant.\label{mtable_rin} 
}
\begin{tabular}{ccccc}
\hline 
& & Suzaku A & Suzaku B & XMM \\
 \hline \hline
 absorption & $N_\mathrm{H} [10^{22}$\,cm$^{-2}]$ & 0.63$_{-0.06}^{+0.07}$ & 0.51$_{-0.08}^{+0.07}$ & 2.32$_{-0.06}^{+0.07}$\\
continuum \& & $A_\text{pex}$ [$10^{-3}$] &  12.82$_{-2.51}^{+0.17}$ & 15.55$_{-0.11}^{+0.15}$ & 12.30$_{-0.15}^{+0.11}$ \\
cold reflection & $\Gamma$ & 1.81$_{-0.01}^{+0.01}$ & 1.92$_{-0.01}^{+0.01}$ & 2.37$_{-0.01}^{+0.02}$ \\
 & $R$ & 0.41$_{-0.03}^{+0.03}$ & 0.40$_{-0.06}^{+0.09}$ & 1.83$_{-0.21}^{+0.13}$\\
 \hline
ionized reflection &$A_\text{reflionx}$ [$10^{-4}$] & 0.31$_{-0.07}^{+0.03}$ & 1.52$_{0.41}^{+0.22}$ & 27.18$_{-4.44}^{+1.33}$\\
 & $Fe/Solar$ & \multicolumn{3}{c}{\textbf{0.81}$_{-0.03}^{+0.01}$} \\
& $\xi$ [erg\,cm\,s$^{-1}$] & 1998.9$_{-135.8}^{+563.5}$ & 14.2$_{-6.1}^{+6.1}$ & 196.78$_{-8.07}^{+5.33}$\\
\hline
relativistic blurring & $q$ & 10.0$_{-8.0}^{+0.0}$ & 2.4$_{-0.2}^{+0.2}$ & 6.97$_{-0.21}^{+0.09}$\\
  &$a$ & \multicolumn{3}{c}{\textbf{0.966}$_{-0.003}^{+0.003}$}\\
 & $i$ [deg] & \multicolumn{3}{c}{\textbf{5}$_{-0}^{+4}$}\\
 & $R_\mathrm{in}$ & 84.3$_{-42.4}^{+14.2}$ & 1$_{-0.0}^{+2.0}$ & 1$_{-0.0}^{+0.1}$ \\
 \hline
plasma & $A_\text{phot}$ & \multicolumn{3}{c}{\textbf{4.6e-5}$_{-7.4e-06}^{+7.96e-06}$}\\
 &$\log\xi$ & \multicolumn{3}{c}{\textbf{1.0}$_{-0.005}^{+0.03}$} \\ 
\hline
individual & $\chi^2$/dof & 1007.4/942 & 1199.6/1403 & 1647.2/1521 \\
 & $\chi^2$/dof & \multicolumn{3}{c}{3854.3/3872} \\
 & $\chi^2_\text{red}$ & \multicolumn{3}{c}{1.00} \\
 \hline\hline
\end{tabular}
\end{center}
\end{table*}

With these hints, we then allowed the inner radius of the accretion disk to vary across epochs (with the constraint that it is no smaller than the ISCO).  Such a fit yields a significantly better goodness-of-fit ($\chi^2/{\rm dof=3854/3872}$ [1.00]; $\Delta \chi^2=214.6$) while at the same time leading to more consistent results. The inclination is now limited to $i<7$\,degrees, more in line with the radio determination of $i<16$\,degrees. The inner radius clearly shows a flux state dependence with it being further out as the X-ray flux becomes smaller. However, it is surprising that the changes in the inner disk radius and disk ionization are of such a magnitude for the two \textit{Suzaku} pointings considering the fact that they are separated by less than a week. The photon indices also show a trend with flux: the higher the flux the steeper the X-ray spectrum. Again, the contribution of ionized disk reflection to the X-ray spectrum increases from 8\,\% to 50\,\% based on the unabsorbed model fluxes in 0.7-10\,keV. The determined spin parameter is high ($a>0.96$). Similar to the fit with fixed inner radius, the iron abundance is sub-solar. The results are consistent with \citet{cowperthwaite:12a}, who for the earlier \textit{Suzaku} pointings (also at time of rising flux) found a possibly recessed disk with high black hole spin. It is curious that the fit requires a significantly variable neutral absorption column, with an especially heavy column during the {\it XMM-Newton} pointing that absorbs a particularly strong soft excess originating from the ionized blurred disk reflection.  See \S\ref{dis} for more discussion.   

\subsubsection{Jet model}\label{jet_m}

\begin{table*}[t]
\begin{center}
\caption{Spectral Parameters for 3C120 multi-epoch fit of the jet model; see the text for a description of the model. \textit{Suzaku} Spectra are normalized to XIS0 data. The power law normalization is $\text{photons}\,\text{keV}^{-1}\,\text{cm}^{-2}\,\text{s}^{-1}$ at 1\,keV. Bold values are the ones which are constrained by all observations as they are assumed to be constant.\label{mtable_alt} 
}
\begin{tabular}{ccccc}
\hline 
& & Suzaku A & Suzaku B & XMM \\
 \hline \hline
 absorption & $N_\mathrm{H} [10^{22}$\,cm$^{-2}]$ & 0.23$_{-0.08}^{+0.08}$ & 0.25$_{-0.11}^{+0.10}$ & 0.28$_{-0.02}^{+0.02}$ \\
continuum \& & $A_\text{pex}$ [$10^{-3}$] &  16.20$_{-0.56}^{+0.60}$ & 17.01$_{-0.67}^{+0.76}$ & 15.31$_{-0.46}^{+0.45}$ \\
cold reflection & $\Gamma$ & 1.96$_{-0.03}^{+0.03}$ & 2.01$_{-0.04}^{+0.04}$ & 1.97$_{-0.03}^{+0.03}$\\
 & $R$ & 1.13$_{-0.15}^{+0.16}$ & 1.17$_{-0.19}^{+0.21}$ & 1.29$_{-0.16}^{+0.17}$\\
 & $Fe/Solar$ & \multicolumn{3}{c}{\textbf{0.28}$_{-0.03}^{+0.03}$} \\
\hline
   soft excess & $\Gamma_S$ & 6.14$_{-0.38}^{+1.09}$ & 5.59$_{-0.33}^{+0.61}$ & 4.83$_{-0.19}^{+0.18}$ \\
   & $A_\text{pow}$ [$10^{-3}$] & 2.86$_{-1.88}^{+2.49}$ & 4.99$_{-3.42}^{+3.49}$ & 7.15$_{-0.60}^{+0.58}$ \\
 \hline
plasma & $A_\text{phot}$ & \multicolumn{3}{c}{\textbf{0.05}$_{-0.04}^{+3.24}$}
\\
 &$\log\xi$ & \multicolumn{3}{c}{\textbf{4.2}$_{-0.0}^{+0.6}$} \\ 
\hline
individual & $\chi^2$/dof & 988.1/946 & 1194.2/1406 & 1651.4/1523 \\
 & $\chi^2$/dof & \multicolumn{3}{c}{3834.5/3881} \\
 & $\chi^2_\text{red}$ & \multicolumn{3}{c}{0.99} \\
 \hline\hline
\end{tabular}
\end{center}
\end{table*}

A multi-epoch fit of the jet model gives $\chi^2/{\rm dof}=3835/3881$ (0.99), a slightly better quality fit than obtained for the variable-$R_{\rm in}$ disk-reflection model.  For the resulting spectral parameters (Table~\ref{mtable_alt}), we find that they are consistent with constant power law  ($\Gamma=1.97$) and reflection fraction ($R\approx1.15$). In addition to this stable hard component the absorption in the Galaxy is also stable at about $2.5\times10^{21}$\,cm$^{-2}$. The only variable part in the spectrum is the soft power law which changes from a slope of $\Gamma_S=6.14_{-0.38}^{+1.09}$ to $\Gamma_S=4.83_{-0.19}^{+0.18}$. The normalization at 1\,keV also approximately doubles towards higher fluxes.   This model does not include a disk reflection component.   Within the context of this model, the ``broad iron line" feature is described by a combination of the iron-K band ``wedge" in the distant reflection component and the photoionized emission.

There are appealing as well as problematic aspects of this spectral solution.   On the positive side, the intrinsic absorption is consistent with being constant ($N_H\approx 2.8\times 10^{21}\,{\rm cm}^{-2}$) as is the reflection fraction ($R\approx 1.15$).  More worryingly, the extremely steep soft powerlaw ($\Gamma_S\sim 5-7$) represents a fine-tuning problem.  This is a steep component that strongly diverges just below our observed low-energy cutoff (0.7\,keV for XIS and 0.5\,keV for EPIC-pn).  Although such an interpretation is not supported by the SED decomposition of \citet{kataoka:11a}, one possible interpretation of such a steep continuum component is that we are catching the extreme (exponentially cut-off) end of the jet synchrotron spectrum.  The question then becomes why this component cuts off at precisely this energy --- a slightly lower cut-off energy would render it invisible in our X-ray spectrum whereas a high cut-off energy would cause it to completely dominate our soft X-ray spectra.   Another troublesome issue is the low iron abundance implied by this fit ($Z\approx 0.28Z_\odot$).   It would be very surprising if the gas in the central regions of this evolved galaxy were not enriched to at least solar levels.  Indeed, in many luminous Seyfert galaxies, central engine abundance enhancements to several times solar are inferred \citep{warner:04a,nagao:06a}.

\subsubsection{Jet$+$disk model}\label{jetdisk}

\begin{table*}[th]
\begin{center}
\caption{Spectral Parameters for 3C120 multi-epoch fit with the jet$+$disk model; see the text for a description of the model. \textit{Suzaku} Spectra are normalized to XIS0 data. The power law normalization is $\text{photons}\,\text{keV}^{-1}\,\text{cm}^{-2}\,\text{s}^{-1}$ at 1\,keV. Bold values are the ones which are constrained by all observations as they are assumed to be constant.\label{mtable_combo} 
}
\begin{tabular}{ccccc}
\hline 
& & Suzaku A & Suzaku B & XMM \\
 \hline \hline
 absorption & $N_\mathrm{H} [10^{22}$\,cm$^{-2}]$ & 0.14$_{-0.02}^{+0.02}$ & 0.18$_{-0.03}^{+0.05}$ & 0.28$_{-0.01}^{+0.01}$ \\
continuum \& & $A_\text{pex}$ [$10^{-3}$] &  10.16$_{-1.24}^{+0.89}$ & 13.19$_{-4.98}^{+0.62}$ & 7.25$_{-0.88}^{+1.48}$ \\
cold reflection & $\Gamma$ & 1.70$_{-0.20}^{+0.75}$ & 1.81$_{-0.04}^{+0.03}$ & 2.40$_{-0.02}^{+0.01}$ \\
 & $R$ & 0.26$_{-0.04}^{+0.03}$ & 0.25$_{-0.06}^{+0.04}$ & 2.00$_{-0.23}^{+0.00}$ \\
 \hline
ionized reflection &$A_\text{reflionx}$ [$10^{-4}$] & 0.76$_{-0.16}^{+0.19}$ & 0.59$_{-0.20}^{+0.27}$ & 58.39$_{-7.41}^{+7.48}$\\
 & $Fe/Solar$ & \multicolumn{3}{c}{\textbf{1.67}$_{-0.08}^{+0.01}$} \\
& $\xi$ [erg\,cm\,s$^{-1}$] & 199.7$_{-14.4}^{+9.8}$ & 123.5$_{-52.7}^{+40.9}$ & 199.0$_{-20.1}^{+8.3}$ \\
\hline
relativistic blurring & $q$ & $>3.5$ & $>2.8$ & 7.8$_{-0.2}^{+0.3}$ \\
  &$a$ & \multicolumn{3}{c}{\textbf{0.994}$_{-0.003}^{+0.002}$}\\
 & $i$ [deg] & \multicolumn{3}{c}{\textbf{5}$_{-0}^{+13}$}\\
 & $R_\mathrm{in}$ & 38.1$_{-20.4}^{+16.8}$ & 39.5$_{-26.2}^{+32.1}$ & 1.0$_{-0.0}^{+0.1}$	 \\
 \hline
   soft excess & $\Gamma_S$ & 2.64$_{-0.14}^{+0.11}$ & 3.80$_{-0.27}^{+0.37}$ & 3.56$_{-0.21}^{+0.14}$ \\
   & $A_\text{pow}$ [$10^{-3}$] & 5.47$_{-1.09}^{+0.45}$ & 5.89$_{-0.79}^{+1.18}$ & 6.23$_{-0.83}^{+0.04}$ \\
 \hline
plasma & $A_\text{phot}$ & \multicolumn{3}{c}{\textbf{6.4e-5}$_{-1.0e-05}^{+1.03e-05}$}\\
 &$\log\xi$ & \multicolumn{3}{c}{\textbf{1.0}$_{-0.0}^{+0.04}$} \\ 
\hline
individual & $\chi^2$/dof & 991.6/940 & 1185.0/1401 & 1625.8/1519 \\
 & $\chi^2$/dof & \multicolumn{3}{c}{3802.4/3867} \\
 & $\chi^2_\text{red}$ & \multicolumn{3}{c}{0.98} \\
 \hline\hline
\end{tabular}
\end{center}
\end{table*}

\begin{figure*}[th]
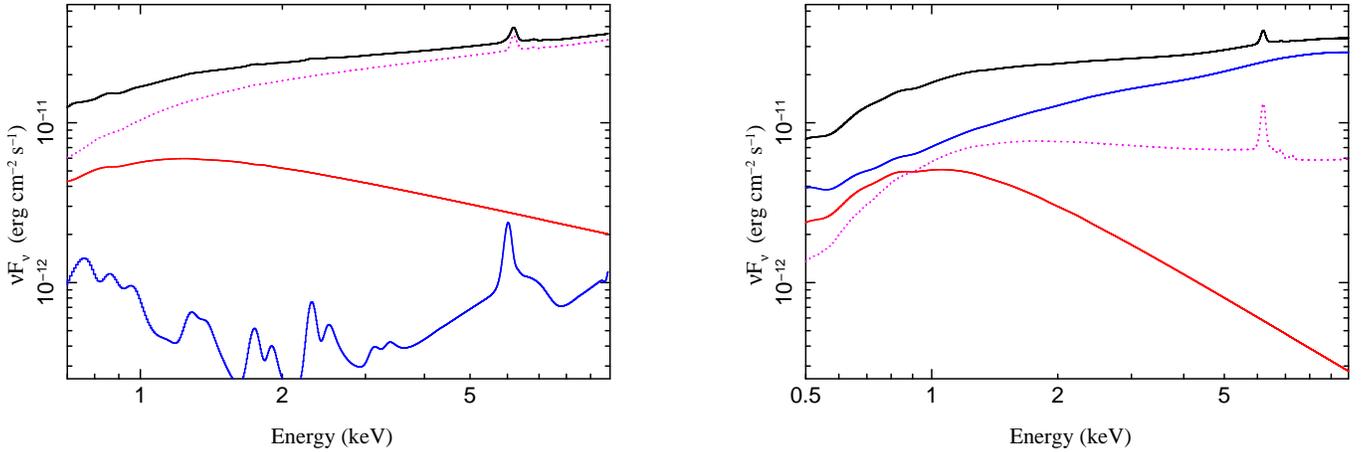

\begin{minipage}{0.45\textwidth}
\includegraphics[width=\columnwidth]{fig6a.ps}
\end{minipage}
\hfill
\begin{minipage}{0.45\textwidth}
\includegraphics[width=\columnwidth]{fig6b.ps}
\end{minipage}
\caption{Examples for the spectral decomposition in the jet+disk model as found to be the best fit for the first \textit{Suzaku} observation (left) and the \textit{XMM} observation (right). Shown here is the total model (solid-thick black line), the primary power-law continuum and cold reflection (dotted magenta line), jet component (solid red line) and the relativistically blurred ionized reflection component (solid blue line).}\label{jetdiskflux}
\end{figure*}

Driven by the potential deficiencies of the ``pure-bred" spectral models discussed above, we explore a mixed model in which we include both a jet component (i.e. an additional steep powerlaw continuum) and reflection from a relativistic disk.  As must be the case (since this model-space completely encompasses the previous models), the goodness of fit parameter shows improvement, $\chi^2/{\rm dof}=3802/3867$ (0.98); this improves on the jet-model by $\Delta\chi^2=32$ (for 14 additional parameters), and the disk-reflection model by $\Delta\chi^2=52$ (for 6 additional parameters). An example of how the different components of this model are able to describe the data is shown in Fig.~\ref{jetdiskflux} for the first \textit{Suzaku} pointing as well as the \textit{XMM} pointing. We note that the broadening of the disk spectrum in case of \textit{XMM} is so severe that the exclusion of the narrow 1.7-2.3\,keV band from our \textit{Suzaku} spectra does not strongly affect a possible detection of such a disk component.    

Especially when comparing the jet model with the jet$+$disk model, an F-test shows that the improvement in fit is not statistically significant.  However, this mixed model solves most of the fine-tuning and physical-inconsistencies that have been noted for the pure-bred models.  In this spectral solution, the jet component has a more modest photon index ($\Gamma_S\sim 2.5-4.0$), alleviating the fine tuning problem noted in \S\ref{jet_m}.  The disk component fits to an almost face-on inclination ($i<18$\,degree) and the implied ionization state of the disk is approximately constant, alleviating the problems with the pure disk-reflection model.  In particular, the large jump in the ionization state of the disk between {\it Suzaku}-A and {\it Suzaku}-B that was seen in the disk-reflection model is no longer present.  This is straightforward to reconcile.  As is evident in Fig.~\ref{basic}, there is a subtle but significant change in the soft excess between these two pointings.  In the disk-reflection model, the soft excess is modeled as blurred ionized reflection and so this spectral change drives a strong change in the inferred disk ionization (even though the rest of the spectrum is very similar).  In the mixed model, the soft excess change is entirely described by a change in the slope of the soft (jet) powerlaw, allowing the accretion disk parameters to remain almost unchanged.  

More noteworthy is the fact that, in this mixed model fit, the intrinsic neutral absorption approximately halves (from $2.8\times 10^{21}\,{\rm cm}^{-2}$ to $1.5\times 10^{21}\,{\rm cm}^{-2}$) between the {\it XMM-Newton} observation in 2003 to the two {\it Suzaku} pointings in 2012.   If we equate this nine year timescale with the dynamical time in the black hole potential ($t_{\rm dyn}=(R^3/GM)^{1/2}$), this implies that the variable absorbing structures are at $10^4R_g$, suggestive of the optical broad line region (BLR).

In this scenario, the {\it XMM-Newton} spectrum is formally reflection dominated, with the ionized reflection from the inner accretion disk dominating the primary continuum across most of the band. At the same time, the inner edge of the accretion disk is strongly constrained to be at the ISCO, the black hole is constrained to be rapidly spinning, and the emissivity index $q$ becomes extremely steep ($q\approx 7.8$).  All of this behavior is consistent with the extreme light bending of a primary X-ray source on the spin axis and very close to a rapidly spinning black hole \citep{miniutti:04a}.  

\section{Short-Term X-ray Spectral Variability}\label{var}

\begin{figure}[!ht]
\includegraphics[width=\columnwidth]{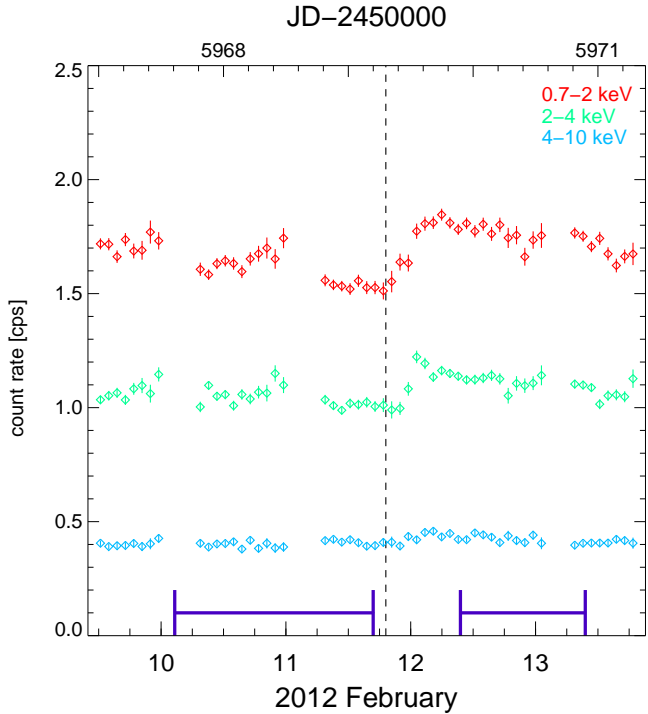}
\caption{Zoom into parts of the XIS3 light curve of the the first Suzaku pointing in different energy bands. Blue lines represent the times from which the low and high state spectrum where constructed. The dashed line marks the approximate onset of the count rate jump in 0.7-2\,keV.}\label{jump}
\end{figure}

After investigating the global long-term variability using long-term light curves (\S \ref{context}) and multi-epoch fitting (\S \ref{spectral}), it is important to note that there is shorter timescale (intra-observation) variability. 
The only significant short-timescale variability observed during the pointings considered in this paper was found in the first {\it Suzaku} pointing, around the beginning of 12-Feb-2012 (Fig~\ref{jump}), in the form of a sudden increase in the count rates below 4\,keV.

\begin{figure}
\includegraphics[width=\columnwidth]{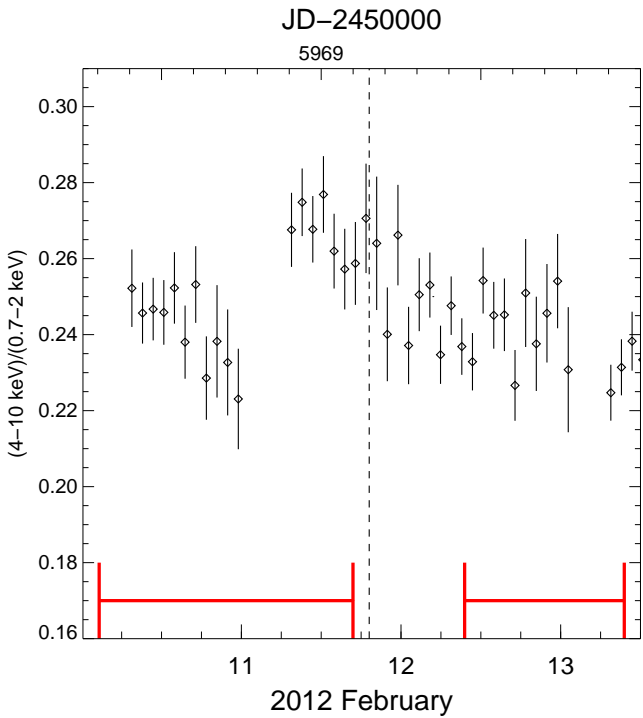}
\caption{Hardness evolution ((4-10\,keV)/(0.7-2\,keV)) during the count rate jump in the \textit{Suzaku} I pointing. The red lines represent the times from which the low and high state spectrum where constructed. The dashed line marks the approximate onset of the count rate jump in 0.7-2\,keV.}\label{hard_jump}
\end{figure}

\begin{figure}[ht]
\includegraphics[width=\columnwidth]{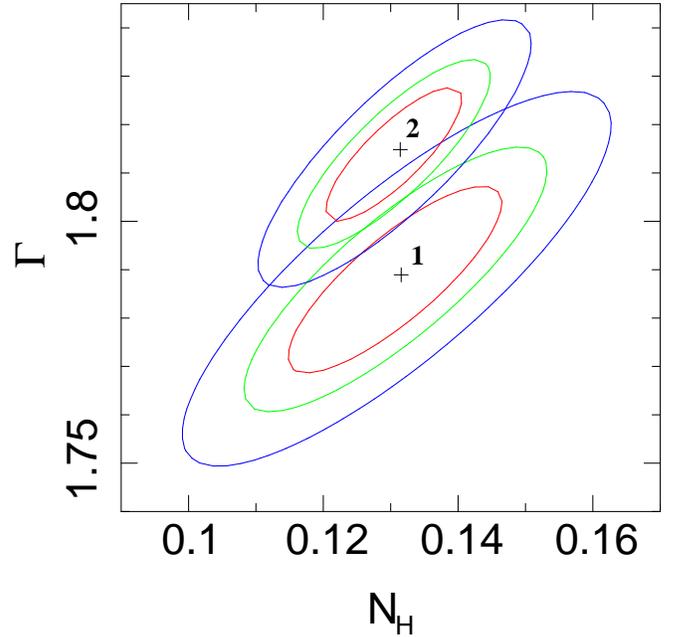}
\caption{Contours of $\Gamma$ and $N_H$ (in units of $10^{22}\,{\rm cm}^{-2}$) for the pre-jump (1) and post-jump (2) spectra of the \textit{Suzaku} I pointing.}\label{contr}
\end{figure}

To explore the nature of this variability we examine the development of the 4--10\,keV/0.7--2\,keV hardness ratio (Fig.~\ref{hard_jump}). We see clear indications for spectral changes, starting even before the sudden count rate increase. A significant hardening of the source is observed in the time preceding the count rate jump. An examination of Fig~\ref{jump} shows that the 4--10\,keV flux jumps before the 2--4\,keV flux which, in turn, peaks before the 0.7--2\,keV band.  
This increase of the 4-10\,keV count rate before the increase of the softer bands explains the observed hardening.  Two possibilities for this sudden hardening and subsequent softening come to mind.  Either the absorption column suddenly increases and then progressively decreases (successively uncovering softer energies) or there is a real change in the continuum shape (with the soft energies lagging the hard band).  To investigate the nature of the variability in more detail, we construct pre-jump and post-jump spectra. The portions of the light curve used for pre-jump and post-jump are indicated in Fig.~\ref{jump}. The spectra can be described by an absorbed power law with a gaussian line as neutral iron K. The quality of the data does not allow us to constrain the absorption and the spectral slope to a degree where we could know which of the two is changing. However, varying only one of the two options we find that adjusting only the absorption leads to a worse fit than allowing for a change in photon index. This is also apparent from the 
contour plot (Fig.~\ref{contr}) of $N_\text{H}$ versus $\Gamma$ which suggests that the continuum slope is varying.

As discussed in \S\ref{me_fitting}, acceptable spectral models attribute at least some of the soft excess to a steep powerlaw-like component from a jet. Thus, while difficult to prove from this one event, we may be witnessing the propagation of a disturbance from the inner accretion disk into the jet.  We discuss this possibility further in \S\ref{dis}.

\section{Discussion}\label{dis}

\subsection{Evidence for a Disk-Jet Connection in 3C120?}\label{diskjet}

Our preferred spectral model to describe the multi-epoch {\it XMM-Newton} and {\it Suzaku} data is the jet$+$disk model (see discussion in \S\ref{jetdisk}).  The fits of this model to the data directly support the predictions of the jet-cycle model.   We find a recessed accretion disk ($R_{\rm in}\approx 20-70R_{\rm isco}$) during the 2012-{\it Suzaku} observations when the X-ray flux is on a rising trend and the jet-cycle picture would suggest a refilling accretion disk.  This scenario is supported by a rising trend in the UV flux.  It is likely that most of UV/optical flux in 3C120 is coming from the disk as the measured polarization is low \citep{chatterjee:09a}, so again the UV rise suggests a refilling accretion disk.  On the other hand, we find that the disk extends all of the way to the ISCO and is strongly irradiated by a compact source close to the black hole during the {\it XMM-Newton} observation when the X-ray flux is at a peak.   To fit this {\it XMM-Newton} spectrum, we require a rapidly rotating black hole.  The formal limits indicate a truly extreme spin $a>0.991$.  However, our methodology implicitly assumes a razor-thin accretion disk (this assumption enters via the termination condition for the ray-tracing underlying the {\tt relconv} relativistic transfer function, see \citealt{dauser:10a}).  For such rapid spins, the ISCO is very close to the event horizon and finite disk-thickness effects must come into play.  While more work is needed on this issue, a first attempt to explore the systematic errors introduced by finite-thickness effects was made by \citet{reynolds:08a} (see their Fig.~5).  This suggests that the formal result of $a>0.991$ may weaken to $a\gtrsim 0.95$ once finite disk thickness and emission within the ISCO is considered.  

Our overall picture is supported by the appearance in the jet of a new superluminal knot in the VLBA images about one month after the 2012-{\it Suzaku} pointings and two months following the middle of the
X-ray dip between JD 2455915 (December, 19th 2011) and 2455960 (February, 2nd 2012). The latter delay is similar to the mean value of $0.18\pm 0.14$ yr observed over five years of {\it RXTE} and VLBA monitoring of 3C120 \citep{chatterjee:09a}. \citet{marscher:02a} and \citet{chatterjee:09a} interpret the delay as the travel time of a disturbance in the jet flow from the base of the jet near the accretion disk to the 43 GHz ``core'' about 0.5 pc away. The enhanced emission 0.2-0.5 mas from the core --- and  $\sim 1$ pc from the central engine --- starting in 2012 August is too bright to be caused by ``trailing shocks" \citep{agudo:01} behind the superluminal knot. A more likely explanation is either the establishment or strengthening of multiple standing shocks. Such shocks are set up by pressure imbalances with the external medium \citep{daly:88a}, which can occur when the energy density injected into the jet varies significantly with time \citep{gomez:97a}. \citet{jorstad:05a} detected such a stationary feature, designated {\it A1},  0.2\,milliarcsec from the core in 3C120 between 1998 and 2001. Figure~\ref{vlba} marks the location of {\it A1}, which is still present in 2012. The brightening of the core, which could be the first in a series of standing shocks, suggests that the rate of injection of energy did indeed increase starting in July 2011. Another apparently stationary emission feature appeared after this about 0.4\,milliarcsec from the core. Unfortunately, our X-ray, UV, and optical observations ended several months before this occurred, hence we cannot determine whether a variation in the disk and corona emission heralded this change in the jet.

\begin{figure}[th]
\includegraphics[width=\columnwidth]{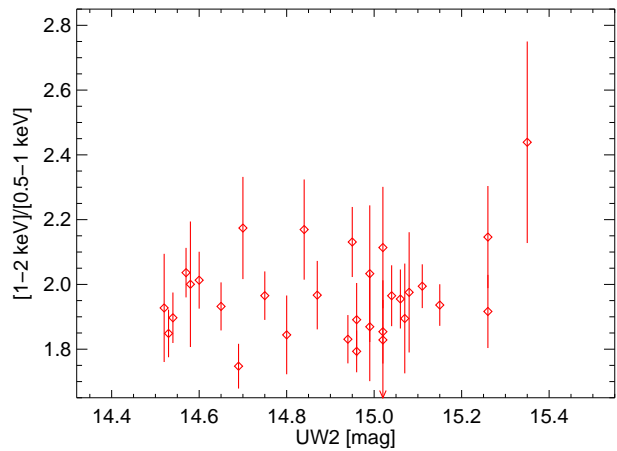}
\caption{Soft X-ray hardness ((1-2\,keV)/(0.3-1\,keV)) versus UV magnitude in the UVW2 band. No strong correlation is apparent between the two.}\label{hardness}
\end{figure}

In our preferred X-ray spectral model, the soft excess is a combination of ionized reflection from the inner disk and a steep ($\Gamma=2.5-4$) jet continuum.  Within the context of standard jet emission models, such steep continua suggest that we are seeing the cut-off region of the jet synchrotron spectrum.  Using the {\it XSPEC} model {\tt srcut}, we have verified that our X-ray spectra would not have been sensitive to the curvature of this synchrotron spectrum since the jet dominates a narrow band at the lowest accessible X-ray energies.   Extrapolating the synchrotron model to lower energies reveals that, in $\nu F(\nu)$ terms, the jet component should be approximately one order of magnitude brighter in the optical-UV band than the soft X-ray.    This is inconsistent with the SED of \citet{kataoka:11a}; however, this SED was constructed from non-simultaneous data and hence must be viewed with caution.  A simultaneous optical/UV/X-ray SED can be constructed from {\it SWIFT} UVOT+XRT data.  \citet{vasudevan:09a} present exactly such a {\it SWIFT} SED, showing indeed that the optical/UV flux exceeds the X-ray flux by a factor of $10--20$.  The existence of a broad-band spectral component extending from the optical/UV into the soft X-ray band is supported by the detection of a correlation between the UV/optical and soft X-ray flux within the \textit{XMM} pointing \citep{ogle:05a}. Similarly, a weak correlation between the (1320\,\AA-1420\,\AA) band and the soft X-ray band (0.3-2\,keV) has also been found previously by \citet{maraschi:91a} from simultaneous IUE-\textit{Exosat} observations. This UV-soft X-ray correlation makes a Comptonization model for the soft excess similar to the one recently discovered in bright Seyferts \citep{lohfink:12b,petrucci:12a} also a possibility in 3C120, while at the same time ruling out a solely reflection-based soft excess. In our UV monitoring, we do not find a correlation between the UV flux and the 1-2\,keV/0.3-1\,keV X-ray hardness ratio (Fig.~\ref{hardness}), as it would be expected from a broad-band component. This absence could be explained in two ways: Firstly, a correlation on the timescales of the monitoring (one week) is not necessarily expected. Secondly, the average \textit{XMM-OM} UVW1 (2910\,\AA) flux was $6.6\times 10^{-14}\,$W\,m$^{-2}$, while our maximum  UVOT-U flux (3465\,\AA) is only $2.9\times 10^{-14}\,$W\,m$^{-2}$ implying that the average flux values must be much lower. If part of this UV/optical flux is in fact a broad-band component its flux contribution could be too low during our monitoring to be detectable. 

\subsection{Variability}

While the spectral analysis enables us to study the spectral components at any given time, variability studies can provide valuable additional information and put constraints on emission regions and the structure of the central accretion flow. Previous analysis of the long-term \textit{RXTE} light curve by \citet{chatterjee:09a} showed that its PSD is similar to radio-quiet AGN, implying the origin of the X-rays may be similar. In agreement with this hypothesis, the spectrum softens as the source brightens \citep{ogle:05a,chatterjee:09a}, as is well established for Seyfert galaxies \citep[e.g., see][]{chiang:00a}.  

While this is all well-known, the considered time scales are rather long, making changes in the accretion rate a likely origin. However the \textit{Swift} and \textit{RXTE} monitorings also indicate short-time scale variability with a time scale of less than a few days.  For example, we find a $35\%$ decrease in the 2--10\,keV count-rate between 1-Dec-2011 and 4-Dec-2011.  However, we find even faster variability in the first of the {\it Suzaku} pointings.  The sudden increase of flux (on a time scale of 12\,hours) starts in the hard band and propagates into the soft-band.  Let us examine the classical time scales characterizing the accretion disk.   For a black hole mass of $\sim6\times10^7\,\rm{M}_\odot$ the dynamical time scale is $t_\text{dyn}\sim \Omega^{-1} \sim {r^{3/2}}/{(G\,M)^{1/2}}\sim$\,2.6\,hrs at $10\,R_g$.  Changes in the local structure of the accretion disk are governed by the thermal time scale which is $t_\text{th}\sim {t_\text{dyn}}/{\alpha}\sim 1\,\text{day}$ at $10\,R_g$ if we assume a disk viscosity parameter of $\alpha=0.1$.  Accretion rate changes occur on the viscous time scale, $t_{\rm visc}\sim t_{\rm dyn}/[\alpha(h/R)^2]$ where $h$ is the disk thickness.  If the radiatively-efficient part of the disk really truncates at $R>20R_g$ during this pointing, as suggested by our preferred spectral fit, the flow at $10R_g$ would be a hot, advection-dominated, geometrically-thick structure ($h\sim R$) implying that $t_{\rm visc}\sim t_\text{th}\sim 1$\,day.  In the light of these time scales, we can see that the short time scale {\it Suzaku} event must have originated from the advection dominated flow close to the black hole (within $10R_g$).  Furthermore, given that the onset of the event in a given band is very rapid ($<6$\,hours), it seems unlikely that the event is driven by changes in accretion rate (viscous time scale) or local disk structure (thermal time scale).  

As already alluded to in \S\ref{var}, the {\it Suzaku} event may be revealing a short-term connection between the accretion disk and the jet.  We will discuss this in the context of our preferred jet$+$disk spectral model.  The {\it Suzaku} event starts as a jump in the hard (4--10\,keV) band that, in our spectral model, is associated with X-ray emission from the disk corona.  Over the next 0.5\,day, it then propagates to the soft (0.7--2\,keV) band that, in our model, has a significant jet component.   If this is the correct interpretation, it gives us a rare look at the short timescale coupling between the disk and the jet.  This coupling is almost certainly magnetic in nature \citep{blandford:77a,blandford:82a}, and introduces a new timescale into the system --- the magnetic timescale.  The magnetic timescale is the timescale on which the poloidal magnetic fields in the disk can spontaneously align in portions of the disk, potentially changing the dissipation in the disk and its coupling to the jet. This magnetic timescale is hard to estimate (ultimately relating to the dynamo problem) but could be very fast \citep{livio:03a,king:04a}, especially in geometrically-thick flows expected within $R_{\rm in}$.  

\section{Summary}

In this paper, we present a detailed spectral analysis of deep {\it XMM-Newton} and {\it Suzaku} pointings of the BLRG 3C~120.   These observations are placed into the context of the hypothesized jet-cycle through the use of {\it RXTE}, {\it Swift} and VLBA monitoring campaigns.  Our main findings are:
\begin{enumerate}
\item Using the monitoring data and the jet-cycle picture, we expect that the 2003-{\it XMM-Newton} spectrum (taken at a time of peak X-ray flux) should show a complete accretion disk extending down to the ISCO, whereas the 2012-{\it Suzaku} observations should show truncated, refilling accretion disks.  
\item A multi-epoch analysis of the {\it XMM-Newton} and {\it Suzaku} pointings finds three statistically acceptable spectral models, a disk-reflection model, a jet-model, and a jet$+$disk model.  While they cannot be distinguished on purely statistical grounds, the disk-reflection model strongly violates the radio constraints on the jet/inner-disk inclination and cannot explain the observed UV-soft X-ray flux correlation. At the same time the jet-model suffers a severe fine-tuning problem.  On the other hand, the jet$+$disk model appears physically reasonable in all respects.  
\item Adopting this jet$+$disk model as our preferred solution, we do indeed find truncated disks during the two {\it Suzaku} pointings and a complete accretion disk at the time of the {\it XMM-Newton} observation.  This is exactly in line with expectations from the jet-cycle picture.  The ejection of a new superluminal knot about two months after an X-ray flux
dip further supports this scenario.  
\item We detect a rapid event in our first {\it Suzaku} pointing that starts in the hard band (4--10\,keV) and propagates to the soft spectrum (0.7--2\,keV).  We interpret this as a disturbance which propagates from the inner (advective) disk into the jet on a timescale of $6-12$\,hours, and suggest that this timescale is set by the dynamics of the magnetic field.  
\item Our preferred spectral solution has a rapid black hole spin.  Formally, our spectral fit yields a spin limit of $a>0.991$.  However, for such rapid spins, finite disk thickness effects must be important and will weaken the limit.  Employing the toy-model of \citet{reynolds:08a} to assess the role of these effects suggests a true limit $a\gtrsim 0.95$.  
\end{enumerate}

\noindent Facilities: \facility{\textsl{Suzaku}}, \facility{\textsl{XMM-Newton}}, \facility{\textsl{Swift}}, \facility{\textsl{RXTE}}, \facility{VLBA}

\vspace{0.5cm}

The authors thank the anonymous referee for comments that improved this manuscript. We thank Phil Cowperthwaite, Javier Garc\`{i}a and Abdu Zoghbi for useful discussions throughout the duration of this work.  AML and CSR gratefully acknowledge NASA for support under the Astrophysical Data Analysis Program (ADAP) grant NNX12AE13G. The research at Boston University was supported by NSF grant AST-907893
and NASA grant NNX11AQ03G. The VLBA is operated by the National Radio Astronomy Observatory. The
National Radio Astronomy Observatory is a facility of the National Science
Foundation operated under cooperative agreement by Associated Universities, Inc.. ACF thanks the Royal Society for support.  


\bibliographystyle{jwapjbib}

\begin{thebibliography}{}

\bibitem[\protect\astroncite{{Ackermann} et~al.}{2011}]{ackermann:11a}
{Ackermann}, M., et~al., 2011, \apj, 743, 171

\bibitem[\protect\astroncite{{Agudo} et~al.}{2001}]{agudo:01}
{Agudo}, I., {G{\'o}mez}, et al.  2001, \apjl, 549, L183

\bibitem[\protect\astroncite{{Agudo} et~al.}{2012}]{agudo:12a}
{Agudo}, I., {G{\'o}mez}, J.~L., {Casadio}, C., {Cawthorne}, T.~V., \&
  {Roca-Sogorb}, M.,  2012, \apj, 752, 92

\bibitem[\protect\astroncite{{Arnaud}}{1996}]{arnaud:96a}
{Arnaud}, K.~A.,  1996,
\newblock in Astronomical Data Analysis Software and Systems V, ed.
  {G.~H.~Jacoby \& J.~Barnes}, Vol. 101, 17

\bibitem[\protect\astroncite{{Ballantyne}, {Fabian} \&
  {Iwasawa}}{2004}]{ballantyne:04a}
{Ballantyne}, D.~R., {Fabian}, A.~C., \& {Iwasawa}, K.,  2004, \mnras, 354, 839

\bibitem[\protect\astroncite{{Bardeen} \&
  {Petterson}}{1975}]{bardeen:75a}
{Bardeen}, J.~M., {Petterson}, \& J.~A.,  1975, \apjl, 195, 65


\bibitem[\protect\astroncite{{Blandford} \& {Payne}}{1982}]{blandford:82a}
{Blandford}, R.~D., \& {Payne}, D.~G.,  1982, \mnras, 199, 883

\bibitem[\protect\astroncite{{Blandford} \& {Znajek}}{1977}]{blandford:77a}
{Blandford}, R.~D., \& {Znajek}, R.~L.,  1977, \mnras, 179, 433

\bibitem[\protect\astroncite{{Belloni}}{2010}]{belloni:10a}
{Belloni}, T.~M.,  2010, 
\newblock in Lecture Notes in Physics, Berlin Springer Verlag, 
Vol. 794, 53


\bibitem[\protect\astroncite{{Chatterjee} et~al.}{2011}]{chatterjee:11a}
{Chatterjee}, R., et~al., 2011, \apj, 734, 43

\bibitem[\protect\astroncite{{Chatterjee} et~al.}{2009}]{chatterjee:09a}
{Chatterjee}, R., et~al., 2009, \apj, 704, 1689

\bibitem[\protect\astroncite{{Chiang} et~al.}{2000}]{chiang:00a}
{Chiang}, J., {Reynolds}, C.~S., {Blaes}, O.~M., {Nowak}, M.~A., {Murray}, N.,
  {Madejski}, G., {Marshall}, H.~L., \& {Magdziarz}, P.,  2000, \apj, 528, 292

\bibitem[\protect\astroncite{{Corbel} et~al.}{2013}]{corbel:13a}
{Corbel}, S., {Coriat}, M., {Brocksopp}, C., {Tzioumis}, A.~K., {Fender},
  R.~P., {Tomsick}, J.~A., {Buxton}, M.~M., \& {Bailyn}, C.~D.,  2013, \mnras,
  428, 2500

\bibitem[\protect\astroncite{{Cowperthwaite} \&
  {Reynolds}}{2012}]{cowperthwaite:12a}
{Cowperthwaite}, P.~S., \& {Reynolds}, C.~S.,  2012, \apjl, 752, L21

\bibitem[\protect\astroncite{{Daly} et~al.}{1988}]{daly:88a}
{Daly}, R. A., \& {Marscher}, A. P. 1988, \apjl, 334, 539

\bibitem[\protect\astroncite{{Dauser} et~al.}{2010}]{dauser:10a}
{Dauser}, T., {Wilms}, J., {Reynolds}, C.~S., \& {Brenneman}, L.~W.,  2010,
  \mnras, 409, 1534
  
\bibitem[\protect\astroncite{{Elvis} et~al.}{1989}]{elvis:89a}
{Elvis}, M., {Wilkes}, B.~J., \& {Lockman}, F.~J., 1989,
  \aj, 97, 777
  

\bibitem[\protect\astroncite{{Fabian} et~al.}{1995}]{fabian:95a}
{Fabian}, A.~C., {Nandra}, K., {Reynolds}, C.~S., {Brandt}, W.~N., {Otani}, C.,
  {Tanaka}, Y., {Inoue}, H., \& {Iwasawa}, K.,  1995, \mnras, 277, L11

\bibitem[\protect\astroncite{{Fender} et~al.}{2004}]{fender:04a}
{Fender}, R.~P., {Belloni}, T.~M., \& {Gallo}, E.,  2004, \mnras, 355, 1105


\bibitem[\protect\astroncite{{Fender} et~al.}{2007}]{fender:07a}
{Fender}, R., {Koerding}, E., {Belloni}, T., {Uttley}, P., {McHardy}, I., \&
  {Tzioumis}, T.,  2007, ArXiv e-prints

\bibitem[\protect\astroncite{{Fender} et~al.}{2009}]{fender:09a}
{Fender}, R.~P., {Homan}, J.,\& {Belloni}, T.~M.,  2009, \mnras, 396,1370

\bibitem[\protect\astroncite{{Garc{\'{\i}}a-Lorenzo} et~al.}{2005}]{garcia:05a}
{Garc{\'{\i}}a-Lorenzo}, B., {S{\'a}nchez}, S.~F., {Mediavilla}, E.,
  {Gonz{\'a}lez-Serrano}, J.~I., \& {Christensen}, L.,  2005, \apj, 621, 146

\bibitem[\protect\astroncite{{G\'omez} et~al.}{1997}]{gomez:97a}
{G\'omez}, J. L., {Mart\'{\i}}, J. M., {Marscher}, A. P., {Ib\'a\~nez}, J. M., \&
{Alberdi}, A. 1997, \apjl, 482, L33

\bibitem[\protect\astroncite{{Houck} \& {Denicola}}{2000}]{houck:00a}
{Houck}, J.~C., \& {Denicola}, L.~A.,  2000,
\newblock in Astronomical Data Analysis Software and Systems IX, ed.
  {N.~Manset, C.~Veillet, \& D.~Crabtree}, Vol. 216, 591

\bibitem[\protect\astroncite{{Jorstad} et~al.}{2005}]{jorstad:05a} 
{Jorstad}, S. G., et al. 2005, \aj, 130, 1418

\bibitem[\protect\astroncite{{Jorstad} et~al.}{2010}]{jorstad:10} 
{Jorstad}, S. G., et al. 2010, \apj, 715, 362

\bibitem[\protect\astroncite{{Kataoka} et~al.}{2007}]{kataoka:07a}
{Kataoka}, J., et~al., 2007, \pasj, 59, 279

\bibitem[\protect\astroncite{{Kataoka} et~al.}{2011}]{kataoka:11a}
{Kataoka}, J., et~al., 2011, \apj, 740, 29

\bibitem[\protect\astroncite{{King} et~al.}{2004}]{king:04a}
{King}, A.~R., {Pringle}, J.~E., {West}, R.~G., \& {Livio}, M.,  2004, \mnras,
  348, 111

\bibitem[\protect\astroncite{{Komatsu} et~al.}{2011}]{komatsu:11a}
{Komatsu}, E., et~al., 2011, \apjs, 192, 18

\bibitem[\protect\astroncite{{Livio}, {Pringle} \& {King}}{2003}]{livio:03a}
{Livio}, M., {Pringle}, J.~E., \& {King}, A.~R.,  2003, \apj, 593, 184

\bibitem[\protect\astroncite{{Lohfink} et~al.}{2012}]{lohfink:12b}
{Lohfink}, A.~M., {Reynolds}, C.~S., {Miller}, J.~M., {Brenneman}, L.~W.,
  {Mushotzky}, R.~F., {Nowak}, M.~A., \& {Fabian}, A.~C.,  2012, \apj, 758, 67

\bibitem[\protect\astroncite{{Maraschi} et~al.}{1991}]{maraschi:91a}
{Maraschi}, L., {Chiappetti}, L., {Falomo}, R., {Garilli}, B., {Malkan}, M.,
  {Tagliaferri}, G., {Tanzi}, E.~G., \& {Treves}, A.,  1991, \apj, 368, 138

\bibitem[\protect\astroncite{{Marscher} et~al.}{2002}]{marscher:02a}
{Marscher}, A.~P., {Jorstad}, S.~G., {G{\'o}mez}, J.-L., {Aller}, M.~F.,
  {Ter{\"a}sranta}, H., {Lister}, M.~L., \& {Stirling}, A.~M.,  2002, \nat,
  417, 625

\bibitem[\protect\astroncite{{Marshall} et~al.}{2009}]{marshall:09a}
{Marshall}, K., {Ryle}, W.~T., {Miller}, H.~R., {Marscher}, A.~P., {Jorstad},
  S.~G., {Chicka}, B., \& {McHardy}, I.~M.,  2009, \apj, 696, 601

\bibitem[\protect\astroncite{{McHardy} et~al.}{2006}]{mchardy:06a}
{McHardy}, I.~M., {Koerding}, E., {Knigge}, C., {Uttley}, P., {Fender}, R.~P.,  2006, \nat, 444, 730

\bibitem[\protect\astroncite{{McKinney} \& {Blandford}}{2009}]{mckinney:09a}
{McKinney}, J.~C., \& {Blandford}, R.~D.,  2009, \mnras, 394, L126

\bibitem[\protect\astroncite{{McKinney}, {Tchekhovskoy} \&
  {Blandford}}{2012}]{mckinney:12a}
{McKinney}, J.~C., {Tchekhovskoy}, A., \& {Blandford}, R.~D.,  2012, \mnras,
  423, 3083

\bibitem[\protect\astroncite{{Merloni} et al.}{2003}]{merloni:03a}
{Merloni}, A., {Heinz}, S., {di Matteo}, T., 2003, \mnras, 345, 1057

\bibitem[\protect\astroncite{{Miller-Jones} et al.}{2012}]{millerjones:12a}
{Miller-Jones}, J.~C.~A., {Sivakoff}, G.~R., {Altamirano}, D., {Coriat}, M., {Corbel}, S., {Dhawan}, V., {Krimm}, H.~A., 	{Remillard}, R.~A., {Rupen}, M.~P., {Russell}, D.~M., {Fender}, R.~P., {Heinz}, S., {K{\"o}rding}, E.~G., {Maitra}, D., {Markoff}, S., {Migliari}, S., {Sarazin}, C.~L.,\& {Tudose}, V., 2012, \mnras, 421, 468


\bibitem[\protect\astroncite{{Miniutti} \& {Fabian}}{2004}]{miniutti:04a}
{Miniutti}, G., \& {Fabian}, A.~C.,  2004, \mnras, 349, 1435

\bibitem[\protect\astroncite{{Mirabel} \& {Rodr{\'{\i}}guez}}{1999}]{mirabel:99a}
{Mirabel}, I.~F. and {Rodr{\'{\i}}guez}, L.~F.,  1999, \araa, 37, 409

\bibitem[\protect\astroncite{{Nagao}, {Maiolino} \&
  {Marconi}}{2006}]{nagao:06a}
{Nagao}, T., {Maiolino}, R., \& {Marconi}, A.,  2006, \aap, 459, 85

\bibitem[\protect\astroncite{{Nandra} et~al.}{2007}]{nandra:07a}
{Nandra}, K., {O'Neill}, P.~M., {George}, I.~M., \& {Reeves}, J.~N.,  2007,
  \mnras, 382, 194

\bibitem[\protect\astroncite{{Ogle} et~al.}{2005}]{ogle:05a}
{Ogle}, P.~M., {Davis}, S.~W., {Antonucci}, R.~R.~J., {Colbert}, J.~W.,
  {Malkan}, M.~A., {Page}, M.~J., {Sasseen}, T.~P., \& {Tornikoski}, M.,  2005,
  \apj, 618, 139

\bibitem[\protect\astroncite{{Petrucci} et~al.}{2012}]{petrucci:12a}
{Petrucci}, P.-O., et~al., 2012, ArXiv e-prints

\bibitem[\protect\astroncite{{Pozo Nu{\~n}ez} et~al.}{2012}]{pozo:12a}
{Pozo Nu{\~n}ez}, F., {Ramolla}, M., {Westhues}, C., {Bruckmann}, C., {Haas},
  M., {Chini}, R., {Steenbrugge}, K., \& {Murphy}, M.,  2012, \aap, 545, A84

\bibitem[\protect\astroncite{{Remillard} \& {McClintock}}{2006}]{remillard:06a}
{Remillard}, R.~A. \& {McClintock}, J.~E., 2006, \araa, 44, 49

\bibitem[\protect\astroncite{{Reynolds} et~al.}{2012}]{reynolds:12a}
{Reynolds}, C.~S., {Brenneman}, L.~W., {Lohfink}, A.~M., {Trippe}, M.~L.,
  {Miller}, J.~M., {Fabian}, A.~C., \& {Nowak}, M.~A.,  2012, \apj, 755, 88

\bibitem[\protect\astroncite{{Reynolds} \& {Fabian}}{2008}]{reynolds:08a}
{Reynolds}, C.~S., \& {Fabian}, A.~C.,  2008, \apj, 675, 1048

\bibitem[\protect\astroncite{{Reynolds}, {Garofalo} \&
  {Begelman}}{2006}]{reynolds:06a}
{Reynolds}, C.~S., {Garofalo}, D., \& {Begelman}, M.~C.,  2006, \apj, 651, 1023

\bibitem[\protect\astroncite{{Reynolds} \& {Nowak}}{2003}]{reynolds:03a}
{Reynolds}, C.~S., \& {Nowak}, M.~A.,  2003, \physrep, 377, 389

\bibitem[\protect\astroncite{{Ross} \& {Fabian}}{2005}]{ross:05a}
{Ross}, R.~R., \& {Fabian}, A.~C.,  2005, \mnras, 358, 211

\bibitem[\protect\astroncite{{Tanaka} et~al.}{1995}]{tanaka:95a}
{Tanaka}, Y., et~al., 1995, \nat, 375, 659

\bibitem[\protect\astroncite{{Tombesi} et~al.}{2011}]{tombesi:11a}
{Tombesi}, F., {Sambruna}, R.~M., {Reeves}, J.~N., {Reynolds}, C.~S.,\& {Braito}, V., 2011, \mnras, 418, 89

\bibitem[\protect\astroncite{{Tombesi} et~al.}{2012}]{tombesi:12a}
{Tombesi}, F., {Sambruna}, R.~M., {Marscher}, A.~P., 
	{Jorstad}, S.~G., {Reynolds}, C.~S.,\& {Markowitz}, A., 2012, \mnras, 424, 754
	
\bibitem[\protect\astroncite{{Vasudevan} \& {Fabian}}{2009}]{vasudevan:09a}
{Vasudevan}, R., \& {Fabian}, A.C., 2009, \mnras, 392, 1124
	
\bibitem[\protect\astroncite{{Warner}, {Hamann} \&
  {Dietrich}}{2004}]{warner:04a}
{Warner}, C., {Hamann}, F., \& {Dietrich}, M.,  2004, \apj, 608, 136

\bibitem[\protect\astroncite{{Wilms}, {Allen} \& {McCray}}{2000}]{wilms:00a}
{Wilms}, J., {Allen}, A., \& {McCray}, R.,  2000, \apj, 542, 914

\bibitem[\protect\astroncite{{Wilms} et~al.}{2006}]{wilms:06a}
{Wilms}, J., {Nowak}, M.~A., {Pottschmidt}, K., {Pooley}, G.~G., \& {Fritz},
  S.,  2006, \aap, 447, 245

\end{thebibliography}

\end{document}